\documentclass{article}
\usepackage[utf8]{inputenc}
\usepackage{graphicx}
\usepackage{times}
\usepackage{xcolor}
\usepackage{amsmath}
\usepackage{subfigure}
\usepackage{amssymb}
\usepackage{epstopdf}
\usepackage[numbers,sort&compress]{natbib}
\usepackage{appendix}
\usepackage{multirow}
\newcommand{\be}{\begin{equation}
\newcommand{\ee}{\end{equation}}}
\newcommand{\bea}{\begin{eqnarray}}
\newcommand{\eea}{\end{eqnarray}}
\newcommand{\nn}{\nonumber}

\begin{document}

\title{Conformable Schr\"odinger Equation in D-dimensional space  }

\author{ Eqab.M.Rabei$^{1,2}$, Mohamed Ghaleb Al-Masaeed$^{3}$,  Sami I. Muslih$^{4, 5}$\\, and Dumitru Baleanu$^{6,7}$\\\\
$^1$Physics Department, Faculty of Science, Al al-Bayt University,\\ P.O. Box 130040, Mafraq 25113, Jordan\\
$^2$Faculty of Science Jerash Private University \\
$^3$Ministry of Education, Jordan\\ 
$^4$Al-Azhar University-Gaza\\
$^5$Southern Illinois University,
Carbondale, IL, 62901 ~USA\\
$^6$ Department of Computer Science and Mathematics.\\
Lebanese American University, Beirut,Lebanon\\
$^7$ Institute of Space Sciences, Magurele–Bucharest, Romania\\\\ Email:eqabrabei@gmail.com\\moh.almssaeed@gmail.com\\sami.muslih@siu.edu\\dumitru.baleanu@lau.edu.lb}

\maketitle


\begin{abstract}
In this work, we extend the time-dependent conformable Schr\"odinger equation for a fractional dimensional system of N spatial coordinates to be used as an effective description of anisotropic and confined systems.  A specific example is looked at in free particle conformable Schr\"odinger wave mechanics, particularly in N-Polar coordinates and N-Cartesian coordinates systems.   The quantities of the conformable form are found to be in exact agreement with the corresponding traditional quantities when $\beta=1$
\\

\textit{Keywords:}  Conformable derivative, Schr\"odinger equation, D-dimensional space, Fractional space
\end{abstract}

\section{Introduction}
The concept of fractional dimension was first developed by mathematician Felix Hausdorff in 1918. This idea became crucial, particularly since Mandelbrot's revolutionary discovery of fractal geometry \cite{mandelbrot1983fractal},   where he applied the idea of fractionality to determine the relationships between fractional and integer dimensions using the scale method i.e. 
\be
d^\alpha x = \frac{\pi^\frac{\alpha}{2} |x|^{\alpha-1}}{\Gamma(\frac{\alpha}{2})} dx, 0<\alpha\leq 1. 
\ee

Low-dimensional systems' confinement has been successfully described physically using fractional dimensional space. This strategy, which was first used by He \cite{he1990anisotropy,he1990fractional,he1991excitons}, substitutes an effective space for the actual confining structure, with the non-integer dimension serving as the measure of its confinement or anisotropy. Additionally, many efforts have been made by researchers throughout a variety of science and technological fields \cite{muslih2007fractional,eid2009fractional,muslih2007fractional1,muslih2011solutions,sadallah2009solution,muslih2011schrodinger,tarasov2015elasticity,ahmad2022capturing,khan2023general,doi:10.1142/S0217984923500021}.\\

Numerous physical systems have been studied using methodologies in fractional-dimensional space. The study of critical phenomena (see, for example, \cite{ma1973introduction,fisher1974renormalization}), of fractal structures \cite{mandelbrot1989fractal}, or in modeling semiconductor heterostructure systems \cite{he1991excitons,mathieu1992simple,christol1993fractional,matos1998fractional,reyes2000excitons} usually takes into consideration theoretical schemes dealing with non-integer space dimensionalities.

The fractional dimensionality in the aforementioned schemes Refers to an additional effective environment that is utilized to represent the real system rather than the real space. However, some authors have also considered the possibility that genuine spacetime has a dimension slightly different from four \cite{zeilinger1985measuring,jarlskog1986number,torres1988blackbody,schafer1986bounds}.

The spacetime dimension has really been observed to deviate from four in a very minor way \cite{zeilinger1985measuring,jarlskog1986number,torres1988blackbody,schafer1986bounds}. Nonetheless, the question of whether the dimension of spacetime is an integer or a fractional number is crucial due to its conceptual significance. Furthermore, there could be intriguing implications if spacetime has a dimension other than four. For instance, it is commonly known that, no matter how tiny the deviation, the logarithmic divergences of quantum electrodynamics are eliminated when the value of four is exceeded. 
\cite{weisskopf1939self}. Also, to see The deformation of quantum mechanics in fractional-dimensional space in Ref \cite{matos2001deformation}

One of the most fascinating recent subjects in a range of physical scientific domains is the application of fractional calculus. In 1695,   L'Hospital and Leibniz exchanged letters during which L'Hospital enquired as to what was meant by a non-integer order of derivatives $\frac {d^n f}{dx^n}$ if $n=\frac{1}{2}$ \cite{oldhamFractionalCalculusTheory1974,millerIntroductionFractionalIntegrals1993}. TheRefore, the idea of a non-integer order of derivatives first emerged from those letters.
The definitions of fractional derivatives have been introduced in a variety of ways throughout history, including Riemann-Louville \cite{oldhamFractionalCalculusTheory1974,millerIntroductionFractionalIntegrals1993}, Caputo \cite{caputoLinearModelsDissipation1967}, Grunwald-Latnikov\cite{grunwald1867uber}, Riesz \cite{marcel1949integrale,riesz1939integrale}, Weyl \cite{weyl1917bemerkungen} and Riesz-Caputo \cite{jarad2012riesz}. Since the majority of them are defined by fractional integrals, they all inherit nonlocal properties from integrals.

Khalil et al.\cite{khalil2014new} proposed a new derivative idea called the conformable derivative a few years ago, and they have utilized the definition of the derivative. theRefore, we can consider the conformable derivative as an extension of the original definition of a derivative. 
The conformable derivative (CD)is one of the definitions that describe the non-integer derivative through the limit definition of the derivative \cite{khalil2014new}.\\
\textbf{Definition 2.1.} Given a function $f\in [0,\infty) \to {R}$. The conformable  derivative of $f$ with order $\beta$ is defined by \cite{khalil2014new}

\be
\label{conformable}
T_\beta(f)(t)=\lim_{\epsilon \to 0}\frac{f(t+\epsilon t^{1-\beta})-f(t)}{\epsilon}
\ee
for all $t>0$, $\beta\in (0.1)$.

The conformable derivative is straightforward in the sense that it fulfills the Leibniz and chain principles as well as the general characteristics of the ordinary derivative.
In \cite{abdeljawad2015conformable,atangana2015new}.\\
 \textbf{Properties} \cite{khalil2014new}
 
 1-$T_\beta(af+bg)=aT_\beta(f)+bT_\beta(g) $\quad for all real constant $a,b$ 
 
 2-$T_\beta(f g)=f T_\beta(g)+g T_\beta(f)$
 
 3-$T_\beta(t^p)=pt^{p-\alpha } $ for all $p$
 
 4-$T_\beta(\frac{f}{g})=\frac{g T_\beta(f)-f T_\beta(g)}{g^2}$
 
 5- $T_\beta(c)=0 $ with  $c$ is constant.\\
 In this paper, we adopt $D^\beta f$ to denote the conformable  derivative (CD) of $f$ of order $\beta$, $ T_\beta(f)(t)$.\\
 The Conformable calculus has successful applications in several fields of physics \cite{mozaffari2018investigation,al2021quantization,chung2020effect,al2022extension,al2022analytic,al2021extension,al2022wkb,al-jamel_effect_2022, HAMMAD2021122307, HAMMAD2021122203,chung2021new,albanwa_quantization_2023,rabei2023treatment}     
\section{Conformable  N-dimensional polar coordinates }
In this section, we would like to obtain a compact expression for the conformable scalar Laplacian in N dimensions.
According to the N-dimensional polar coordinates in Ref \cite{erdelyi1953higher,blumenson1960derivation}, moreover the conformable trigonometric function in Ref \cite{chung2020effect}, and conformable spherical coordinates in Ref \cite{al2022extension}. The conformable  N-dimensional polar coordinates read as  
\bea
\nn
x_1^\beta  &=& r^\beta  \cos{\frac{\theta_1^\beta}{\beta}},\\\nn
x_2^\beta &=& r^\beta \sin{\frac{\theta_1^\beta}{\beta}} \cos{ \frac{\theta_2^\beta}{\beta}},\\\nn
x_3^\beta &=& r^\beta \sin{ \frac{\theta_1^\beta}{\beta}}\sin{ \frac{\theta_2^\beta}{\beta}} \cos{ \frac{\theta_3^\beta}{\beta}},\\\nn
&\cdots& \\\nn
x_{N-1}^\beta &=& r^\beta \sin{ \frac{\theta_1^\beta}{\beta}}\sin{ \frac{\theta_2^\beta}{\beta}} \sin{ \frac{\theta_3^\beta}{\beta}} \cdots \sin{ \frac{\theta_{N-2}^\beta}{\beta}}\cos{ \frac{\theta_{N-1}^\beta}{\beta}},\\\nn
x_{N}^\beta &=& r^\beta \sin{ \frac{\theta_1^\beta}{\beta}}\sin{ \frac{\theta_2^\beta}{\beta}} \sin{ \frac{\theta_3^\beta}{\beta}} \cdots \sin{\frac{\theta_{N-1}^\beta }{\beta}}.
\eea
Where $0\leq \theta_a \leq \pi, (a=1,2,3, \cdots, N-2), 0\leq \theta_{N-1} \leq 2\pi $.
\be
\nn
 r^\beta =\sqrt{x_1^{2\beta} + x_2^{2\beta}+ \cdots+x_{N-1}^{2\beta} + x_N^{2\beta}}
\ee
\bea
\frac{\theta_1^\beta}{\beta} &=& \arccos{\frac{x_1}{\sqrt{x_1^{2\beta} + x_2^{2\beta}+ \cdots+x_{N-1}^{2\beta} + x_N^{2\beta}}}},\\\nn
\frac{\theta_2^\beta}{\beta} &=& \arccos{\frac{x_2}{\sqrt{ x_2^{2\beta}+ \cdots+x_{N-1}^{2\beta} + x_N^{2\beta}}}},\\\nn
 \frac{\theta_3^\beta}{\beta}&=& \arccos{\frac{x_3}{\sqrt{x_3^{2\beta}+ \cdots+x_{N-1}^{2\beta} + x_N^{2\beta}}}},\\\nn
&\cdots&\\\nn
 \frac{\theta_{N-2}^\beta}{\beta}&=& \arccos{\frac{x_{N-2}}{\sqrt{x_{N-2}^{2\beta} + x_{N-1}^{2\beta} + x_N^{2\beta}}}},\\\nn
\frac{\theta_{N-1}^\beta }{\beta}&=& \arccos{\frac{x_{N-1}}{\sqrt{x_{N-1}^{2\beta} + x_N^{2\beta}}}}.
\eea
 Let us consider the scalar  Laplacian operator in  N-Dimensional was suggested in Ref \cite{palmer2004equations}, and the conformable gradient  and conformable scalar  Laplacian operator in 3-D were  used in \cite{atangana2015new,al2022analytic}. TheRefore, the  Conformable scalar  Laplacian operator in  N-Dimensional read as   
\be
\label{dell}
\nabla^{2\beta}_D= \sum_{i=1}^N \partial^{2\beta}_{x_i}+ \frac{\alpha_i -1}{\frac{x_i^\beta}{\beta}} \partial_{x_i}^\beta
\ee
Now let us analyze  the conformable Schr\"odinger equation for any N, eq\eqref{dell}.
\bea 
\nn
\nabla^{2\beta}_D\psi_(x_i)&=& \left(\partial^{2\beta}_{x_1} + \beta \frac{\alpha_1 -1}{x_1^\beta} \partial^{\beta}_{x_1}+\partial^{2\beta}_{x_2}+ \beta \frac{\alpha_2 -1}{x_2^\beta} \partial^{\beta}_{x_2}+ \dots+\partial^{2\beta}_{x_N}+ \beta \frac{\alpha_N -1}{x_N^\beta} \partial^{\beta}_{x_N} \right)\psi_(x_i)
\eea
Choosing $\alpha_N$ as the single parameter for the non-integer dimensions with  $\alpha_1=\alpha_2=\cdots=\alpha_{N-1}=1, D_s=\alpha_N+(N-1)$. In this case, equation \eqref{dsh2} can be put in the form
\bea 
\label{nabla2d}
\nabla^{2\beta}_D\psi_(x_i)&=& \left(\partial^{2\beta}_{x_1} +\partial^{2\beta}_{x_2}+ \dots+\partial^{2\beta}_{x_N}+ \beta \frac{\alpha_N -1}{x_N^\beta} \partial^{\beta}_{x_N} \right)\psi_(x_i)
\eea
After making the calculations, we have 
\bea
\nn
\nabla^{2\beta}_D&=&\partial^{2\beta}_{r}+
\frac{\beta(D_s-1)}{r^\beta}\partial^{\beta}_{r}+
\frac{\beta^2}{r^{2\beta}} \left[\partial^{2\beta}_{\theta_1}
+\frac{(D_s-2)}{\tan{\frac{\theta_1^\beta}{\beta}}}\partial^{\beta}_{\theta_1} \right]+
\frac{\beta^2}{r^{2\beta} \sin^2{\frac{\theta_1^\beta}{\beta} }} \left[\partial^{2\beta}_{\theta_2}+
 \frac{(D_s-3)}{ \tan{\frac{\theta_2^\beta}{\beta} }}\partial^{\beta}_{\theta_2}\right]
 \\\nn&+&
\frac{\beta^2} {r^{2\beta}} \frac{ 1 }{\sin^2{\frac{\theta_1^\beta}{\beta} } \sin^2{\frac{\theta_2^\beta}{\beta}} } \left[\partial^{2\beta}_{\theta_3} +\frac{(D_s-4)}{  \tan{ \frac{\theta_3^\beta}{\beta}}} \partial^{\beta}_{\theta_3}\right]
+\dots \\ \label{ncdell}&+& \frac{\beta^2}{r^{2\beta} \sin^2{\frac{\theta_1^\beta}{\beta}} \sin^2{\frac{\theta_2^\beta}{\beta}} \dots \sin^2{\frac{\theta_{N-2}^\beta}{\beta}} } \left( \partial^{2\beta}_{\theta_{N-1}} +
\frac{D_s -N}{\tan{\frac{\theta_{N-1}^\beta}{\beta}}}  \partial^{\beta}_{\theta_{N-1}}\right)
\eea
\section{Conformable Schr\"odinger Equation in N-Dimensional }
In this section, we would like to obtain the Conformable Schr\"odinger Equation in N-Dimensional.
The time-dependent Conformable Schr\"odinger equation in 3-D reads \cite{chung2020effect}
\be
\label{TIME shro}
\left(-\frac{\hbar^{2\beta}_\beta }{2m^\beta}\nabla^{2\beta}+V_\beta(\hat{x}_\beta)\right)\Psi_\beta(x_i,t) = i \hbar^\beta_\beta D^\beta_t \Psi_\beta(x_i,t).
\ee
In a stationary state \cite{chung2020effect} when $\Psi_\beta(x_i,t)= \psi_\beta(x_i) T_\beta(t^\beta)= \psi_\beta(x_i)  e^{-\frac{i}{\hbar^{\beta}_\beta}E^\beta \frac{t^\beta}{\beta}} $.\\ So, 
\be
\label{TIse shro}
\left(\nabla^{2\beta}- \frac{ 2m^\beta}{\hbar^{2\beta}_\beta}(V_\beta(\hat{x}_\beta) - E^\beta )\right)\psi_\beta(x_i) = 0.
\ee
This equation time-independent Conformable Schr\"odinger equation in 3-D \cite{chung2020effect}. To generalization of conformable Schr\"odinger equation to N-Dimensions, we get 
\be 
\label{dsh1}
\left(\nabla^{2\beta}_D- \frac{ 2m^\beta}{\hbar^{2\beta}_\beta}(V_\beta(\hat{x}_\beta) - E^\beta )\right)\psi_\beta(x_i) = 0.
\ee 
After substituting eq.\eqref{ncdell}, we have 

\bea 
\nn
&&\partial^{2\beta}_{r}+
\frac{\beta(D_s-1)}{r^\beta}\partial^{\beta}_{r}+
\frac{\beta^2}{r^{2\beta}} \left[\partial^{2\beta}_{\theta_1}
+\frac{(D_s-2)}{\tan{\frac{\theta_1^\beta}{\beta}}}\partial^{\beta}_{\theta_1} \right]+
\frac{\beta^2}{r^{2\beta} \sin^2{\frac{\theta_1^\beta}{\beta} }} \left[\partial^{2\beta}_{\theta_2}+
 \frac{(D_s-3)}{ \tan{\frac{\theta_2^\beta}{\beta} }}\partial^{\beta}_{\theta_2}\right]
 \\\nn&+&
\frac{\beta^2} {r^{2\beta}} \frac{ 1 }{\sin^2{\frac{\theta_1^\beta}{\beta} } \sin^2{\frac{\theta_2^\beta}{\beta}} } \left[\partial^{2\beta}_{\theta_3} +\frac{(D_s-4)}{  \tan{ \frac{\theta_3^\beta}{\beta}}} \partial^{\beta}_{\theta_3}\right]
+\dots \\ \nn &+& \frac{\beta^2}{r^{2\beta} \sin^2{\frac{\theta_1^\beta}{\beta}} \sin^2{\frac{\theta_2^\beta}{\beta}} \dots \sin^2{\frac{\theta_{N-2}^\beta}{\beta}} } \left( \partial^{2\beta}_{\theta_{N-1}} +
\frac{D_s -N}{\tan{\frac{\theta_{N-1}^\beta}{\beta}}}  \partial^{\beta}_{\theta_{N-1}} \right)
\\ \label{dsh2} &-& \frac{ 2m^\beta}{\hbar^{2\beta}_\beta}(V_\beta(\hat{x}_\beta) - E^\beta )\psi_\beta(x_i) = 0.
\eea
\section{Free particle }
In this section, we would like to illustrate the Conformable Schr\"odinger Equation in N-Dimensional by solving it for Free particles.
\subsection{Free particle in N-dimensional space polar coordinates}
For free particles $V_\beta(\hat{x}_\beta)=0$  the eq.\eqref{dsh2} becomes as 
\bea 
\nn
&&\partial^{2\beta}_{r} \psi_\beta(x_i)+
\frac{\beta(D_s-1)}{r^\beta}\partial^{\beta}_{r} \psi_\beta(x_i) +
\frac{\beta^2}{r^{2\beta}} \left[\partial^{2\beta}_{\theta_1}
+\frac{(D_s-2)}{\tan{\frac{\theta_1^\beta}{\beta}}}\partial^{\beta}_{\theta_1} \right] \psi_\beta(x_i)
\\\nn&+&
\frac{\beta^2}{r^{2\beta} \sin^2{\frac{\theta_1^\beta}{\beta} }} \left[\partial^{2\beta}_{\theta_2}+
 \frac{(D_s-3)}{ \tan{\frac{\theta_2^\beta}{\beta} }}\partial^{\beta}_{\theta_2}\right] \psi_\beta(x_i)
 \\\nn&+&
\frac{\beta^2} {r^{2\beta}} \frac{ 1 }{\sin^2{\frac{\theta_1^\beta}{\beta} } \sin^2{\frac{\theta_2^\beta}{\beta}} } \left[\partial^{2\beta}_{\theta_3} +\frac{(D_s-4)}{  \tan{ \frac{\theta_3^\beta}{\beta}}} \partial^{\beta}_{\theta_3}\right]\psi_\beta(x_i)
+\dots \\ \nn &+& \frac{\beta^2}{r^{2\beta} \sin^2{\frac{\theta_1^\beta}{\beta}} \sin^2{\frac{\theta_2^\beta}{\beta}} \dots \sin^2{\frac{\theta_{N-2}^\beta}{\beta}} } \left( \partial^{2\beta}_{\theta_{N-1}} +
\frac{D_s -N}{\tan{\frac{\theta_{N-1}^\beta}{\beta}}}  \partial^{\beta}_{\theta_{N-1}} \right)\psi_\beta(x_i)
\\ \label{dsh3} &+& \frac{ 2m^\beta}{\hbar^{2\beta}_\beta} E^\beta \psi_\beta(x_i) = 0.
\eea
Using the Separation of functions 
\bea
\psi_\beta(x_i) = R_\beta (r^\beta) \Pi_{i=1}^{N-1} \Theta_{i\beta}(\theta^\beta).
\eea
So, eq.\eqref{dsh3} becomes as 
\bea 
\nn
&&\frac{1}{ R_\beta}\left[\partial^{2\beta}_{r} R_\beta+
\frac{\beta(D_s-1)}{r^\beta}\partial^{\beta}_{r}  R_\beta\right] +
\frac{\beta^2}{r^{2\beta}  \Theta_{1\beta}} \left[\partial^{2\beta}_{\theta_1}
+\frac{(D_s-2)}{\tan{\frac{\theta_1^\beta}{\beta}}}\partial^{\beta}_{\theta_1} \right]  \Theta_{1\beta}
\\\nn&+&
\frac{\beta^2}{r^{2\beta}  \Theta_{2\beta}\sin^2{\frac{\theta_1^\beta}{\beta} }} \left[\partial^{2\beta}_{\theta_2}+
 \frac{(D_s-3)}{ \tan{\frac{\theta_2^\beta}{\beta} }}\partial^{\beta}_{\theta_2}\right]  \Theta_{2\beta}
 \\\nn&+&
\frac{\beta^2} {r^{2\beta}  \Theta_{3\beta}} \frac{ 1 }{\sin^2{\frac{\theta_1^\beta}{\beta} } \sin^2{\frac{\theta_2^\beta}{\beta}} } \left[\partial^{2\beta}_{\theta_3} +\frac{(D_s-4)}{  \tan{ \frac{\theta_3^\beta}{\beta}}} \partial^{\beta}_{\theta_3}\right] \Theta_{3\beta}
+\dots \\ \nn &+& \frac{\beta^2}{r^{2\beta}  \Theta_{(N-1)\beta} \sin^2{\frac{\theta_1^\beta}{\beta}} \sin^2{\frac{\theta_2^\beta}{\beta}} \dots \sin^2{\frac{\theta_{N-2}^\beta}{\beta}} } \left( \partial^{2\beta}_{\theta_{N-1}} +
\frac{D_s -N}{\tan{\frac{\theta_{N-1}^\beta}{\beta}}}  \partial^{\beta}_{\theta_{N-1}} \right)\Theta_{(N-1)\beta}
\\ \label{dsh4} &+& \frac{ 2m^\beta}{\hbar^{2\beta}_\beta} E^\beta \psi_\beta(x_i) = 0.
\eea
Thus, the conformable radial equation is 
\bea
\frac{1}{ R_\beta}\left[\partial^{2\beta}_{r} R_\beta+
\frac{\beta(D_s-1)}{r^\beta}\partial^{\beta}_{r}  R_\beta\right]+ \frac{ 2m^\beta}{\hbar^{2\beta}_\beta} E^\beta  -
\frac{\beta^2 \lambda_r}{r^{2\beta}  } =0.
\eea
The conformable angular part is
\bea
\partial^{2\beta}_{\theta_1}\Theta_{1\beta}
+\frac{(D_s-2)}{\tan{\frac{\theta_1^\beta}{\beta}}}\partial^{\beta}_{\theta_1}   \Theta_{1\beta} &+& \left[\lambda_r-\frac{\lambda_1}{\sin^2{\frac{\theta_1^\beta}{\beta} }}\right]\Theta_{1\beta}=0.
\\
\partial^{2\beta}_{\theta_2}\Theta_{2\beta}
+\frac{(D_s-3)}{\tan{\frac{\theta_2^\beta}{\beta}}}\partial^{\beta}_{\theta_2}   \Theta_{2\beta} &+& \left[\lambda_1-\frac{\lambda_2}{\sin^2{\frac{\theta_2^\beta}{\beta} }}\right]\Theta_{2\beta}=0.
\\\nn &\vdots&\\ \label{n-2}
\partial^{2\beta}_{\theta_{N-2}}\Theta_{\beta({N-2})}
+\frac{(D_s-(N-1))}{\tan{\frac{\theta_{N-2}^\beta}{\beta}}}\partial^{\beta}_{\theta_{N-2}}   \Theta_{\beta({N-2})} &+& \left[\lambda_{N-3}-\frac{\lambda_{N-2}}{\sin^2{\frac{\theta_{N-2}^\beta}{\beta} }}\right]\Theta_{\beta({N-2})}=0.
\\\nn &\vdots&\\ \label{n-1}
\partial^{2\beta}_{\theta_{N-1}}\Theta_{\beta({N-1})}
+\frac{(D_s-N)}{\tan{\frac{\theta_{N-1}^\beta}{\beta}}}\partial^{\beta}_{\theta_{N-1}}   \Theta_{\beta({N-1})} &+&\lambda_{N-2} \Theta_{\beta({N-1})}=0.
\eea
\subsection{The solution of the conformable radial part}
\bea
\label{R1}
\partial^{2\beta}_{r} R_\beta+
\frac{\beta(D_s-1)}{r^\beta}\partial^{\beta}_{r}  R_\beta+ \beta^2 \left[K^2  -
\frac{  \lambda_r}{r^{2\beta}  }\right] R_\beta =0
\eea
where $K^2=\frac{ 2m^\beta}{\hbar^{2\beta}_\beta \beta^2 } E^\beta$. Let $r^\beta =\frac{\rho^\beta}{K} \to \partial^{\beta}_{r} = K \partial^{\beta}_{\rho} \to \partial^{2\beta}_{r} = K^2 \partial^{2\beta}_{\rho} $. Thus, we have 
\bea
\label{R2}
\partial^{2\beta}_{\rho}  R_\beta+
\frac{\beta(D_s-1)}{\rho^\beta} \partial^{\beta}_{\rho}   R_\beta + \beta^2  \left[1 -
\frac{ \lambda_r}{\rho^{2\beta}  }\right] R_\beta =0
\eea
So,
\bea
\label{R3}
\rho^{2\beta} \partial^{2\beta}_{\rho}  R_\beta +
\beta a \rho^\beta \partial^{\beta}_{\rho}   R_\beta + \beta^2 \left[\rho^{2\beta} -
 \lambda_r \right] R_\beta =0
\eea
where $a=(D_s-1)$ and $\lambda_r$ is called separation constant $\lambda_r = \ell(\ell+ D_s -3)$ \cite{stillinger1977axiomatic,he1991excitons}. This equation  is called    Conformable  Bessel’s differential equation \cite{martinez2022novel}. according to the solution for this equation in \cite{martinez2022novel},  we have
\bea
R_\beta =\rho^{\beta \frac{1-a}{2}}\left[ A J_{\beta q (\rho)} + B J_{-\beta q} (\rho)\right],
\eea
where $q= \frac{\sqrt{a^2+1-2a+4 \lambda_r}}{2}$ . After substituting $a=(D_s-1)$ and $\lambda_r = \ell(\ell+ D_s -3)$, we obtain
\bea
\label{Rsol}
R_\beta =\rho^{\beta\left(1-\frac{ D_s}{2}\right)} \left[ A J_{\beta q } (\rho) + B J_{-\beta q} (\rho)\right].
\eea
where $q= \sqrt{\frac{D_s^2}{4} - D_s+1+\ell(\ell+ D_s -3)}$. $J_{\beta q} (\rho)$ it is called  the first solution of the conformable Bessel equation of order $\beta q$
\bea
J_{\beta q} (\rho)=\rho^{\beta\left(1-\frac{ D_s}{2}\right)}  \sum_{s=0}^\infty \frac{(-1)^s }{ s! \Gamma(s+q+1)} \left(\frac{\rho^{\beta}}{2}\right)^{2s+q}.
\eea
And $J_{-\beta q} (\rho)$ it is called  the Second solution of the conformable Bessel equation of order $\beta q$
\bea
J_{-\beta q} (\rho)=\rho^{\beta\left(1-\frac{ D_s}{2}\right)}  \sum_{s=0}^\infty \frac{(-1)^s }{ s! \Gamma(s-q+1)} \left(\frac{\rho^{\beta}}{2}\right)^{2s-q}.
\eea
after substituting $r^\beta =\frac{\rho^\beta}{K}$,we have 
\bea
J_{\beta q} (r K)=  \sum_{s=0}^\infty \frac{(-1)^s }{ s! \Gamma(s+q+1)} \left(\frac{r^{\beta} K}{2}\right)^{2s+q}.
\eea
\bea
J_{-\beta q} (rK)=  \sum_{s=0}^\infty \frac{(-1)^s }{s! \Gamma(s-q+1)}\left(\frac{r^{\beta} K}{2}\right)^{2s-q}.
\eea
So, the solution in eq.\eqref{Rsol}
\bea
\label{Rsol}
R_\beta =(r^\beta K)^{\left(1-\frac{ D_s}{2}\right)} \left[ A J_{\beta q} (rK) + B J_{-\beta q} (rK)\right].
\eea

$B$, the constant, must equal zero since the  Schr\"odinger equation's solution must be finite and regular.
\bea
\label{Rsol12}
R_\beta =(r^\beta K)^{\left(1-\frac{ D_s}{2}\right)}  A J_{\beta q} (rK) .
\eea

\subsection{The solution of the conformable angular part} 
\subsubsection{N-1 term}
We can use the change of variable to solve eq.\eqref{n-1}.
Let
\bea
\nn
\Theta_{\beta({N-1})}&=&X_{\beta({N-1})},\\\nn
\cos{\frac{\theta_{N-1}^\beta}{\beta}} &=&  x_{N-1}^\beta,\\\nn
\partial^{\beta}_{\theta_{N-1}}&=&-\frac{1}{\beta}\sin{\frac{\theta_{N-1}^\beta}{\beta}}\partial^{\beta}_{x_1},\\\nn
\partial^{2\beta}_{\theta_{N-1}}&=&\frac{1}{\beta^2}\sin^2{\frac{\theta_{N-1}^\beta}{\beta}} \partial^{2\beta}_{x_{N-1}},\\\nn
\partial^{2\beta}_{\theta_{N-1}}&=&\frac{(1-x^{2\beta}_{N-1})}{\beta^2} \partial^{2\beta}_{x_{N-1}}.
\eea
So, the  eq.\eqref{n-1} becomes 
\bea
\label{xn-1eq}
(1-x^{2\beta}_{N-1}) \partial^{2\beta}_{x_{N-1}}
X_{\beta({N-1})}-\beta (D_s-N)x_{N-1}^\beta \partial^{\beta}_{\theta_{N-1}} X_{\beta({N-1})}  +\beta^2 \lambda_{N-2} X_{\beta({N-1})}=0.
\eea
When comparing this equation with the Conformable Gegenbauer equation[cite conformable Gegenbauer] 
\bea
(1-x^{2\beta})D_x^\beta D_x^\beta T_{\beta m_{N-3}}^{\gamma_{N-1}}(x)&-&2\beta(\gamma_{N-1}+1) x^\beta D_x^\beta T_{\beta m_{N-3}}^{\gamma_{N-1}}(x) \\\nn&+& \beta^2 m_{N-3}(m_{N-3}+2\gamma_{N-1}+1) T_{\beta m_{N-3}}^{\gamma_{N-1}}(x)=0.
\eea
So, the eigenvalue $\lambda_{N-2}=m_{N-3}(m_{N-3}+D_s-N-1)$ where $\gamma_{N-1}=\frac{D_s-N-2}{2}$ and the solution of eq.\eqref{n-1}  is given by 
\be
X_{\beta({N-1})}=T_{\beta m_{N-3}}^{\gamma_{N-1}}(x).
\ee
Where $T_{\beta m_{N-3}}^{\gamma_{N-1}}(x)$ is the Conformable Gegenbauer polynomials [cite conformable Gegenbauer].\\ 
\subsubsection{N-2 term}
Thus, to solve eq.\eqref{n-2} we use the change of variable  
Let 
\bea
\nn
X_{\beta({N-2})}&=&  \Theta_{\beta({N-2})},\\\nn
x_{N-2}^\beta &=& \cos{\frac{\theta_{N-2}^\beta}{\beta}},\\\nn \partial^{\beta}_{\theta_{N-2}}&=&
-\frac{1}{\beta}\sin{\frac{\theta_{N-2}^\beta}{\beta}}\partial^{\beta}_{x_{N-2}},\\\nn\partial^{2\beta}_{\theta_{N-2}}&=&
\frac{1}{\beta^2}\sin^2{\frac{\theta_{N-2}^\beta}{\beta}} \partial^{2\beta}_{x_{N-2}},\\\nn\partial^{2\beta}_{\theta_{N-2}} &=&
\frac{(1-x^{2\beta}_{N-2})}{\beta^2} \partial^{2\beta}_{x_{N-2}}. 
\eea
So, the  eq.\eqref{n-2} becomes 
\bea
\label{n-21}
(1-x^{2\beta}_{N-2}) \partial^{2\beta}_{x_{N-2}}
X_{\beta({N-2})}&-&\beta (D_s-N+1)\partial^{\beta}_{x_{N-2}}X_{\beta({N-2})}\\ \nn&+& \left[\lambda_{N-3}-\frac{m_{N-3}(m_{N-3}+2\gamma_{N-2})}{1-x^{2\beta}_{N-2}}\right]X_{\beta({N-2})}=0.
\eea
where $\gamma_{N-2}=\gamma_{N-1}+ \frac{1}{2}=\frac{D_s-N-1}{2}$. To calculate the eigenvalue $\lambda_{N-3}$ let $m_{N-3}=0$,so, this equation becomes 
\bea
(1-x^{2\beta}_{N-2}) \partial^{2\beta}_{x_{N-2}}
X_{\beta({N-2})}-\beta (D_s-N+1)\partial^{\beta}_{x_{N-2}}X_{\beta({N-2})}+\lambda_{N-3}X_{\beta({N-2})}=0.
\eea
When comparing this equation with the Conformable Gegenbauer equation[cite conformable Gegenbauer], we obtain 
\bea
\nn
\lambda_{N-3}=m_{N-4}(m_{N-4}+2\gamma_{N-2}+1)=m_{N-4}(m_{N-4}+D_s-N).
\eea

To solve eq.\eqref{n-21} we suppose 
\bea
\label{x111}
X_{\beta({N-2})}= (1-x^{2\beta}_{N-2})^{\frac{m_{N-3}}{2}} u_{(N-2)\beta},
\eea
where 
\bea
\label{dx111}
D^\beta_{x_1} X_{\beta({N-2})}= (1-x^{2\beta})^{\frac{m_{N-3}}{2}} D^\beta_{x_1}u_{(N-2)\beta}-m_{N-3} \beta x^\beta (1-x^{2\beta})^{\frac{m_{N-3}}{2}-1}u_{(N-2)\beta} ,
\eea
and
\bea
D^\beta_{x_{N-2}}D^\beta_{x_{N-2}} X_{\beta({N-2})}&=& (1-x^{2\beta}_{N-2})^{\frac{m_{N-3}}{2}} D^\beta_{x_{N-2}}D^\beta_{x_{N-2}}u_{(N-2)\beta}
\\\nn&-&2m_{N-3} \beta x^\beta_{N-2} (1-x^{2\beta}_{N-2})^{\frac{m_{N-3}}{2}-1}D^\beta_{x_{N-2}}u_{(N-2)\beta}
\\\nn&-&m_{N-3} \beta^2  (1-x^{2\beta}_{N-2})^{\frac{m_{N-3}}{2}-1}u_{(N-2)\beta}
\\\nn&+&m_{N-3}(m_{N-3}-2) \beta^2 x^{2\beta}_{N-2} (1-x^{2\beta}_{N-2})^{\frac{m_{N-3}}{2}-2}u_{(N-2)\beta} ,
\eea 
after multiplying this equation by $(1-x^{2\beta})$, we have 
\bea
\label{ddx111}
(1-x^{2\beta}_{N-2})D^\beta_{x_1}D^\beta_{x_{N-2}} X_{\beta({N-2})}&=& (1-x^{2\beta}_{N-2})^{\frac{m_{N-3}}{2}+1} D^\beta_{x_{N-2}} D^\beta_{x_{N-2}} u_{(N-2)\beta}
\\\nn&-&2m_{N-3} \beta x^\beta_{N-2} (1-x^{2\beta}_{N-2})^{\frac{m_{N-3}}{2}}D^\beta_{x_{N-2}}u_{(N-2)\beta}
\\\nn&-&m_{N-3} \beta^2  (1-x_{N-2}^{2\beta})^{\frac{m_{N-3}}{2}}u_{(N-2)\beta}
\\\nn&+&m_{N-3}(m_{N-3}-2) \beta^2 x^{2\beta}_{N-2} (1-x^{2\beta}_{N-2})^{\frac{m_{N-3}}{2}-1}u_{(N-2)\beta}.
\eea
Thus, after substituting eqs.\eqref{x111},\eqref{dx111}, and \eqref{ddx111} in \eqref{n-21} and  multiplying by $(1-x^{2\beta})^{\frac{m}{2}}$, we have 
\bea
 (1-x^{2\beta}_{N-2}) D^\beta_{x_{N-2}} D^\beta_{x_{N-2}} u_{(N-2)\beta}&-&2m_{N-3} \beta x^\beta_{N-2} D^\beta_{x_{N-2}}u_{(N-2)\beta}
-m_{N-3} \beta^2  u_{(N-2)\beta}
\\\nn&+&m_{N-3}(m_{N-3}-2) \beta^2 x^{2\beta}_{N-2} (1-x^{2\beta}_{N-2})^{-1}u_{(N-2)\beta}\\\nn&-& 2\beta x^\beta_{N-2} (\gamma_{N-2}+1)D^\beta_{x_{N-2}}u_{(N-2)\beta}
\\\nn&+&2\beta (\gamma_{N-2}+1)m_{N-3} \beta x^{2\beta}_{N-2} (1-x^{2\beta}_{N-2})^{-1}u_{(N-2)\beta}\\\nn&+&  \beta^2 \left[m_{N-4}(m_{N-4}+2\gamma_{N-2}+1)-\frac{m_{N-3}(m_{N-3}+2\gamma_{N-2})}{1-x^{2\beta}_{N-2}}\right]u_{(N-2)\beta}
\eea
so, we have 
\bea
\nn
&&(1-x^{2\beta}_{N-2}) D_{x_{N-2}}^\beta D_{x_{N-2}}^\beta u_{(N-2)\beta}-2\beta x^\beta_{N-2}(m_{N-4} + \gamma_{N-2}+1) D^\beta_{x_{N-2}}u_{(N-2)\beta}
\\ \label{x1122}&+&  \beta^2 [m_{N-4}(m_{N-4}+2\gamma_{N-2}+1)- m_{N-3}(m_{N-3}+2 \gamma_{N-2} +1 )]u_{(N-2)\beta}=0.
\eea 
The Conformable Gegenbauer equation[cite conformable Gegenbauer]
\bea
(1-x^{2\beta}_{N-2})D_{x_{N-2}}^\beta D_{x_{N-2}}^\beta T_{\beta m_{N-4}}^{\gamma_{N-2}}(x)&-&2\beta(\gamma_{N-2}+1) x_{N-2}^\beta D_{x_{N-2}}^\beta T_{\beta m_{N-4}}^{\gamma_{N-2}}(x) 
\\\nn&+& \beta^2 m_{N-4}(m_{N-4}+2\gamma_{N-2}+1) T_{\beta m_{N-4}}^{\gamma_{N-2}}(x)=0.
\eea
Taking the conformable derivative of order $m_{N-3}$, 
\bea
\nn
D^{m_{N-3} \beta}[(1-x^{2\beta}_{N-2})D_{x_{N-2}}^\beta D_{x_{N-2}}^\beta T_{\beta m_{N-4}}^{\gamma_{N-2}}(x)]&-&2\beta(\gamma_{N-2}+1)  D^{m_{N-3} \beta}[ x_{N-2}^\beta D_{x_{N-2}}^\beta T_{\beta m_{N-4}}^{\gamma_{N-2}}(x)] \\\nn&+& \beta^2 m_{N-4}(m_{N-4}+2\gamma_{N-2}+1) T_{\beta m_{N-4}}^{\gamma_{N-2}}(x)\\\label{CGEG2}&=&0.
\eea
after using conformable Leibniz rule\cite{rabei_solution_2021}, we have
\bea
\label{dm22}
D^{m_{N-3} \beta}[(1-x^{2\beta}_{N-2}) D_{x_{N-2}}^\beta D_{x_{N-2}}^\beta T_{\beta m_{N-4}}^{\gamma_{N-2}}(x)]&=&(1-x^{2\beta}_{N-2}) D^{m_{N-3} \beta} D_x^\beta D_x^\beta T_{\beta m_{N-4}}^{\gamma_{N-2}}(x)
\\\nn&-&2m_{N-3}\beta x^\beta_{N-2} D^{m_{N-3} \beta}D_x^\beta T_{\beta m_{N-4}}^{\gamma_{N-2}}(x)\\ \nn&-&m_{N-3}(m_{N-3}-1) \beta^2 D^{m_{N-3} \beta}T_{\beta m_{N-4}}^{\gamma_{N-2}}(x)
\eea
\bea
\label{dm21}
D^{m_{N-3} \beta}[ x_{N-2}^\beta D_{x_{N-2}}^\beta T_{\beta m_{N-4}}^{\gamma_{N-2}}(x)]&=&x^{\beta}_{N-2} D^{m_{N-3} \beta}  D^\beta_{x_{N-2}} T_{\beta m_{N-4}}^{\gamma_{N-2}}(x)
\\\nn&-&m_{N-3}\beta D^{m_{N-3} \beta} T_{\beta m_{N-4}}^{\gamma_{N-2}}(x)
\eea
Substituting eqs\eqref{dm22}and \eqref{dm21} in \eqref{CGEG2}, we have
\bea
\nn
(1-x^{2\beta}_{N-2}) D_{x_{N-2}}^\beta D_{x_{N-2}}^\beta D^{m_{N-3} \beta} T_{\beta m_{N-4}}^{\gamma_{N-2}}(x)&-&2\beta x^\beta_{N-2}(m_{N-4} + \gamma_1+1) D^\beta_{x_{N-2}}D^{m_{N-4} \beta} T_{\beta m_{N-4}}^{\gamma_{N-2}}(x)
\\\nn +  \beta^2 [m_{N-4}(m_{N-4}+2\gamma_{N-2}+1)&-& m_{N-3}(m_{N-3}+2 \gamma_{N-2} +1 )]D^{m_{N-4} \beta}T_{\beta m_{N-4}}^{\gamma_{N-2}}(x)=0.
\eea 
comparing this equation with eq.\eqref{x1122} we find the eq.\eqref{x111} becomes 
\be
X_{\beta({N-2})}= (1-x^{2\beta}_{N-2})^{\frac{m_{N-3}}{2}} D^{m_{N-3} \beta} T_{\beta m_{N-4}}^{\gamma_{N-2}}(x)
\ee

the separations constants take the values \cite{palmer2004equations}, 
\bea
\lambda_r &=& \ell(\ell+D-3),\quad \ell=0,1,\cdots,
\\ \lambda_1 &=& m_0(m_0+D-4), \quad m_0=0,1, \cdots, m_0 \leq \ell, \\
\lambda_2 &=& m_1(m_1+D-5), \quad m_1=0,1, \cdots, m_1 \leq m_0
\\&\vdots&\\
\lambda_{N-3} &=& m_{N-4}(m_{N-4}+D-N),  m_{N-4}=0,1, \cdots, m_{N-4} \leq m_{N-5},\\
\lambda_{N-2} &=& m_{N-3}(m_{N-3}+D-N-1), \quad m_{N-3}=0,1, \cdots, m_{N-4} \leq m_{N-5}.
\eea
The generalized quantum number $n,  \ell ,m_0, m_1, m_2,..., m_{N-2}$, clearly follows the following quantization rule \cite{sadallah2009solution}

\be
m_{N-3} \leq m_{N-4} \leq m_{N-5} \cdots  \leq m_0  \leq \ell \leq n
\ee
\subsection{Free particle in N-dimensional space Cartesian  coordinates}
The time-independent conformable Schr\"odinger equation in N-dimensional space Cartesian  coordinates through using eqs.\eqref{nabla2d} and \eqref{dsh1} becomes 
\be 
\nn
\left(\partial^{2\beta}_{x_1} +\partial^{2\beta}_{x_2}+ \dots+\partial^{2\beta}_{x_N}+ \beta \frac{\alpha_N -1}{x_N^\beta} \partial^{\beta}_{x_N}- \frac{ 2m^\beta}{\hbar^{2\beta}_\beta}(V_\beta(\hat{x}_\beta) - E^\beta )\right)\psi_\beta(x_i) = 0.
\ee 
In free particle after using $ D_s=\alpha_N+(N-1)$ this equation becomes 
\be 
\left(\partial^{2\beta}_{x_1} +\partial^{2\beta}_{x_2}+ \dots+\partial^{2\beta}_{x_N}+ \beta \frac{D_s-N }{x_N^\beta} \partial^{\beta}_{x_N}+ \frac{ 2m^\beta}{\hbar^{2\beta}_\beta}  E^\beta\right)\psi_\beta(x_i) = 0.
\ee 
Using the separation of functions $\psi_\beta(x_i)=\Pi_{i=1}^{N} X_{i\beta}(x^\beta) $, we obtain 
\bea
\nn
\frac{1}{X_{1\beta}}\partial^{2\beta}_{x_1} X_{1\beta} &+&\frac{1}{X_{2\beta}} \partial^{2\beta}_{x_2} X_{2\beta}+ \dots +\frac{1}{X_{N\beta}}\left[ \partial^{2\beta}_{x_N}X_{N\beta}+ \beta \frac{D_s-N}{x_N^\beta} \partial^{\beta}_{x_N}X_{N\beta} \right] \\&+& \frac{ 2m^\beta}{\hbar^{2\beta}_\beta} ( E^\beta_{x_1}+E^\beta_{x_2}+\dots+E^\beta_{x_N})  = 0.
\eea 
So, we have 
\bea
\partial^{2\beta}_{x_1} X_{1\beta} &+&K_{x_1}^2 X_{1\beta}=0, \quad K_{x_1}^2 = \frac{ 2m^\beta}{\hbar^{2\beta}_\beta}E^\beta_{x_1}.\\
\partial^{2\beta}_{x_2} X_{2\beta} &+& K_{x_2}^2 X_{2\beta}=0, \quad K_{x_2}^2 = \frac{ 2m^\beta}{\hbar^{2\beta}_\beta}E^\beta_{x_2}.\\\nn
&\vdots&\\ \label{Nddcar}
\partial^{2\beta}_{x_N}X_{N\beta}&+& \beta \frac{D_s-N}{x_N^\beta} \partial^{\beta}_{x_N}X_{N\beta} +K_{x_N}^2 X_{N\beta}=0, \quad K_{x_N}^2 = \frac{ 2m^\beta}{\hbar^{2\beta}_\beta}E^\beta_{x_N}
\eea
Thus, the solution of these equations are  
\bea
X_{1\beta}&=& A_{x_1} \exp{\left(ik_{x_1} \frac{x_1^\beta}{\beta}\right)}\\
X_{1\beta}&=& A_{x_2} \exp{\left(ik_{x_2} \frac{x_2^\beta}{\beta}\right)}\\&\vdots\\
X_{(N-1)\beta}&=& A_{x_{N-1}} \exp{\left(ik_{x_{N-1}} \frac{x_{N-1}^\beta}{\beta}\right)}.
\eea
And the solution of eq.\eqref{Nddcar}
\bea
\partial^{2\beta}_{x_N}X_{N\beta}&+& \beta \frac{D_s-N}{x_N^\beta} \partial^{\beta}_{x_N}X_{N\beta} +K_{x_N}^2 X_{N\beta}=0, 
\eea
after multiplying this equation by $x_{N}^{2\beta}$, we have
\bea
x_{N}^{2\beta}\partial^{2\beta}_{x_N}X_{N\beta}+ \beta a x_{N}^{\beta}\partial^{\beta}_{x_N}X_{N\beta} +k_{x_N}^2 x_{N}^{2\beta}X_{N\beta}=0, 
\eea
Let $x_{N}^\beta =\frac{\rho^\beta}{k_{x_N}} \to \partial^{\beta}_{x_{N}} = k_{x_N} \partial^{\beta}_{\rho} \to \partial^{2\beta}_{x_{N}} = k_{x_N}^2 \partial^{2\beta}_{\rho} $. Thus, we have 
\bea
\rho^{2\beta}\partial^{2\beta}_{\rho} X_{N\beta}+ \beta a \rho^\beta \partial^{\beta}_{\rho} X_{N\beta}+ \rho^{2\beta}X_{N\beta}=0, 
\eea
where $a=(D_s-N)$. This equation is similar to eq.\eqref{R3} when $\lambda_r=0$. So, the solution for this equation is given by  
\bea
X_{N\beta} =\rho^{\beta \frac{1-a}{2}}\left[ A J_{\beta q }(\rho) + B J_{-\beta q} (\rho)\right],
\eea
where $q= \frac{1-a}{2}$ . After substituting $a=(D_s-N),x_{N}^\beta =\frac{\rho^\beta}{k_{x_N}}$, we obtain
\bea
X_{N\beta} =x_{N}^{\beta q } k_{x_N}^{\beta q }\left[ A J_{\beta q }(x_{N}k_{x_N}) + B J_{-\beta q} (x_{N}k_{x_N})\right],
\eea
where $q= \frac{1-D_s+N}{2}$. The solution of the Schr\"odinger equation must be finite and regular, so, the constant $B$ must be equal to zero.
\bea
X_{N\beta} =x_{N}^{\beta q } k_{x_N}^{\beta q } A J_{\beta q }(x_{N}k_{x_N}) ,
\eea
The total wave function for free particles in N-dimensional space is 
\bea
\nn
\psi_\beta(x_i)&=&\Pi_{i=1}^{N} X_{i\beta}(x^\beta)=X_{1\beta}(x^\beta)X_{2\beta}(x^\beta)\dots X_{N\beta}(x^\beta)\\&=&C x_{N}^{\beta q } J_{\beta q }(x_{N}k_{x_N}) \Pi_{j=1}^{N-1}  \exp{\left(ik_{x_j} \cdot \frac{x_j^\beta}{\beta}\right)}   
\eea
where $k_{x_j} \cdot \frac{x_j^\beta}{\beta}=k_{x_1} \frac{x_1^\beta}{\beta}+k_{x_2} \frac{x_2^\beta}{\beta}+\cdots+k_{x_N} \frac{x_N^\beta}{\beta}$.
The total energy for free particle in N-dimensional space is 
\bea
\nn
E^\beta &=&  E^\beta_{x_1}+E^\beta_{x_2}+\dots+E^\beta_{x_N}\\&=&\frac{ \hbar^{2\beta}_\beta}{2m^\beta}[k_{x_1}^2+k_{x_2}^2+\dots+k_{x_N}^2].
\eea
\subsection{For 3D}
$N=D_s=3\to q=\frac{1}{2}$. So, we have 
\bea
\nn
\psi_\beta(x_i)&=&Cx_{3}^{\beta  \frac{1}{2} } J_{\beta \frac{1}{2} }(x_{3}k_{x_3}) \exp{\left(ik_{x_1} \frac{x_1^\beta}{\beta}\right)}   \exp{\left(ik_{x_2} \frac{x_2^\beta}{\beta}\right)} 
\eea
Where in Ref  \cite{gökdoğan2015conformable}, $J_{\beta \frac{1}{2} }(x_{3}k_{x_3})=\sqrt{\frac{2}{\pi x^\beta}} \sin\left(k_{x_3}\frac{x^\beta_3 }{\beta}\right)$, we have 
\bea
\nn
\psi_\beta(x_i)&=&C\sqrt{\frac{2}{\pi }} \sin\left(k_{x_3}\frac{x^\beta_3 }{\beta}\right) \exp{\left(ik_{x_1} \frac{x_1^\beta}{\beta}\right)}   \exp{\left(ik_{x_2} \frac{x_2^\beta}{\beta}\right)} 
\eea

\section{Conclusions}
The time-dependent conformable Schr\"odinger equation For a fractional dimensional system of N-spatial coordinates is extended to be applicable for the anisotropic and confined system. The time-independent conformable Schr\"odinger equation is generalized to N-dimensions and it is solved for a free particle. The separation method is used to obtain the solution. The conformable Bessel function is obtained as a solution of the radial part. In addition, the solution of the angular part is obtained. After some manipulation the conformable Gegenbauer equation is recovered, moreover, the generalized quantum numbers are obtained from the separation constants. Besides, In section (4.4), the Schr\"odinger equation for a fractional dimensional system of N special coordinates for free particles in Cartesian coordinates is constructed and solved, The total energy for free particles in N-dimensional space is obtained, and we observed one may recover the traditional solution for 3D.
\bibliography{Ref} 
\bibliographystyle{apalike}
\end{document}